\documentclass[twocolumn,showpacs,preprintnumbers,amsmath,aps]{revtex4}
\usepackage{graphicx}
\usepackage{dcolumn}
\usepackage{bm}
\begin{document}
\newcommand{\kvec}{\mbox{{\scriptsize {\bf k}}}}
\def\eq#1{(\ref{#1})}
\def\fig#1{\hspace{1mm}\ref{#1}}
\def\tab#1{table\hspace{1mm}\ref{#1}}
\title{
---------------------------------------------------------------------------------------------------------------\\
THE PROPERTIES OF THE SUPERCONDUCTING STATE IN $\rm{YNi_{2}B_{2}C}$: THE ONE-BAND ELIASHBERG APPROACH}
\author{M.W. Jarosik, R. Szcz{\c{e}}{\`s}niak}
\affiliation{1. Institute of Physics, Cz{\c{e}}stochowa University of Technology, Al. Armii Krajowej 19, 42-200 
                Cz{\c{e}}stochowa, Poland}
\email{jarosikmw@wip.pcz.pl}
\date{\today} 
\begin{abstract}
The basic thermodynamic parameters of the superconducting state in $\rm{YNi_{2}B_{2}C}$ were calculated in the framework of the one-band Eliashberg model. The effective Eliashberg function, determined on the basis of the transport function (R.S. Gonnelli, {\it et al.}, Physica C \textbf{341}, 1957 (2000)), was used during calculations. It was shown that the dimensionless ratios are equal to: 
$R_{1}\equiv 2\Delta\left(0\right)/k_{B}T_{C}=3.87$, $R_{2}\equiv\Delta C\left(T_{C}\right)/C^{N}\left(T_{C}\right)=1.79$ and 
$R_{3}\equiv T_{C}C^{N}\left(T_{C}\right)/H_{C}^{2}\left(0\right)=0.159$. The value $R_{1}$ fairly agrees with the experimental data whereas $R_{2}$ and $R_{3}$ agree very well.
\end{abstract}
\pacs{74.20.Fg, 74.25.Bt, 74.25.Ha}
\maketitle
%
\section{Introduction}

The superconducting state in the most of the low temperature superconductors is being induced by the electron-phonon interaction \cite{Carbotte}, \cite{Carbotte1}. The coupling between the electron and phonon system can be modeled with a use of the Fr\"{o}hlich Hamiltonian \cite{Frohlich}. Let us notice that using a canonical transformation, during the elimination of the phonon degrees of freedom in Fr\"{o}hlich operator, it is possible to obtain the Hamiltonian of the BCS theory \cite{TransKanon}, \cite{BCS}. It should be clearly marked that the BCS model is able to accurately describe the thermodynamic properties of the low temperature superconductors in the limit of the weak coupling, i.e. for $\lambda<0.2$, where $\lambda$ denotes the electron-phonon coupling constant.

In order to precisely estimate the thermodynamic parameters of the superconductors that have a greater value of $\lambda$, the approach proposed by Eliashberg should be used \cite{Eliashberg}. In the Eliashberg scheme the analysis of the superconductivity issue is being started directly from the Fr\"{o}hlich Hamiltonian, which is initially written in the Nambu notation \cite{Nambu}. Next, with a use of the matrix Matsubara functions the Dyson equations are determined; the self-energy of the system is being calculated with an accuracy to the second order in the equations of motion \cite{Carbotte}, \cite{Carbotte1}, \cite{Eliashberg}. In the last step, the Eliashberg set is determined in a self-consistent way.

The set of the Eliashberg equations generalizes the fundamental equation of the BCS model. In particular, one can take into consideration the complicated form of the electron-phonon interaction with a use of the Eliashberg function. Moreover, the application of the full version of the self-consistent method enables to estimate the electron band effective mass in the presence of the electron-phonon interaction and the energy shift function that renormalizes the electron band energy \cite{Carbotte}.

From the mathematical point of view the solution of the Eliashberg set is a truly complicated matter. For that reason, the effect of the electron band energy renormalization is usually omitted what results in the reduction of the equation's number by one third. The approximation given above, in the most cases, does not affect the final results in a significant manner. This is evidenced by the good agreement between theoretical predictions and the experimental data \cite{Szczesniak}. It need to be marked that the strict solution of the simplified Eliashberg set is not easy and can be performed only with a use of a powerful computer and highly advanced numerical methods \cite{Szczesniak1}.

The electron-phonon coupling constant for $\rm{YNi_{2}B_{2}C}$ superconductor is equal to $0.676$ \cite{Jarosik}. In this case, the exact estimation of the thermodynamic parameters is possible only in the framework of the Eliashberg approach. In the literature related to the topic there are some reports that the simplest version of the Eliashberg equations might not be able to cope with the description of all relevant physical quantities. In particular, the dependence of the upper critical field ($H_{C2}$) on the temperature \cite{Shulga}, \cite{Doh} or the results obtained with a use of the directional point-contact spectroscopy \cite{Muhhopadhyay}, \cite{Bashlakov} suggest the necessity of the two-band model being used. On the other hand, some researchers basing on the thermal and spectroscopic experiments are moving toward the direction of the one-band models with a non-trivial wave symmetry ($s+g$ or even $d$-wave symmetry) \cite{Maki}-\cite{Nohara2}. The issue is clouding by the fact that the calculations conducted till the present day for the one-band Eliashberg model were not strict; the study were based on very simplified form of the Eliashberg function \cite{Gonnelli} or approximated analytical formulas \cite{Michor}. 

For this reson, in the presented paper, the most important thermodynamic parameters for $\rm{YNi_{2}B_{2}C}$ superconductor were analyzed exactly in the framework of the one-band Eliashberg model. The calculations were based on the effective Eliashberg function:  
$\left[\alpha^{2}F\left(\Omega\right)\right]_{\rm{eff}}=1.283 \left[\alpha^{2}F\left(\Omega\right)\right]_{\rm{tr}}$, where the transport function 
($\left[\alpha^{2}F\left(\Omega\right)\right]_{\rm{tr}}$) was determined by Gonnelli {\it et al.} in \cite{Gonnelli} (for details see also \cite{Jarosik}).
  
\section{THE ELIASHBERG EQUATIONS}

The Eliashberg equations on the imaginary axis can be written in the following form \cite{Eliashberg}:
\begin{equation}
\label{r1}
\Delta_{n}Z_{n}=\frac{\pi}{\beta} \sum_{m=-M}^{M}
\frac{K\left(\omega_{n}-\omega_{m}\right)} 
{\sqrt{\omega_m^2+\Delta_m^2}}
\Delta_{m}, 
\end{equation}
\begin{equation}
\label{r2}
Z_n=1+\frac {\pi}{\beta\omega _n }\sum_{m=-M}^{M}
\frac{K\left(\omega_{n}-\omega_{m}\right)}
{\sqrt{\omega_m^2+\Delta_m^2}}\omega_m,
\end{equation}
where $\Delta_{n}\equiv\Delta\left(i\omega_{n}\right)$ represents the order parameter and 
$Z_{n}\equiv Z\left(i\omega_{n}\right)$ is the wave function renormalization factor; the $n$-th Matsubara frequency is denoted by: $\omega_{n}\equiv \frac{\pi}{\beta}\left(2n-1\right)$, where $\beta\equiv 1/k_{B}T$ ($k_{B}$ is the Boltzmann constant).
     
\begin{figure} [ht]
\includegraphics[scale=0.15]{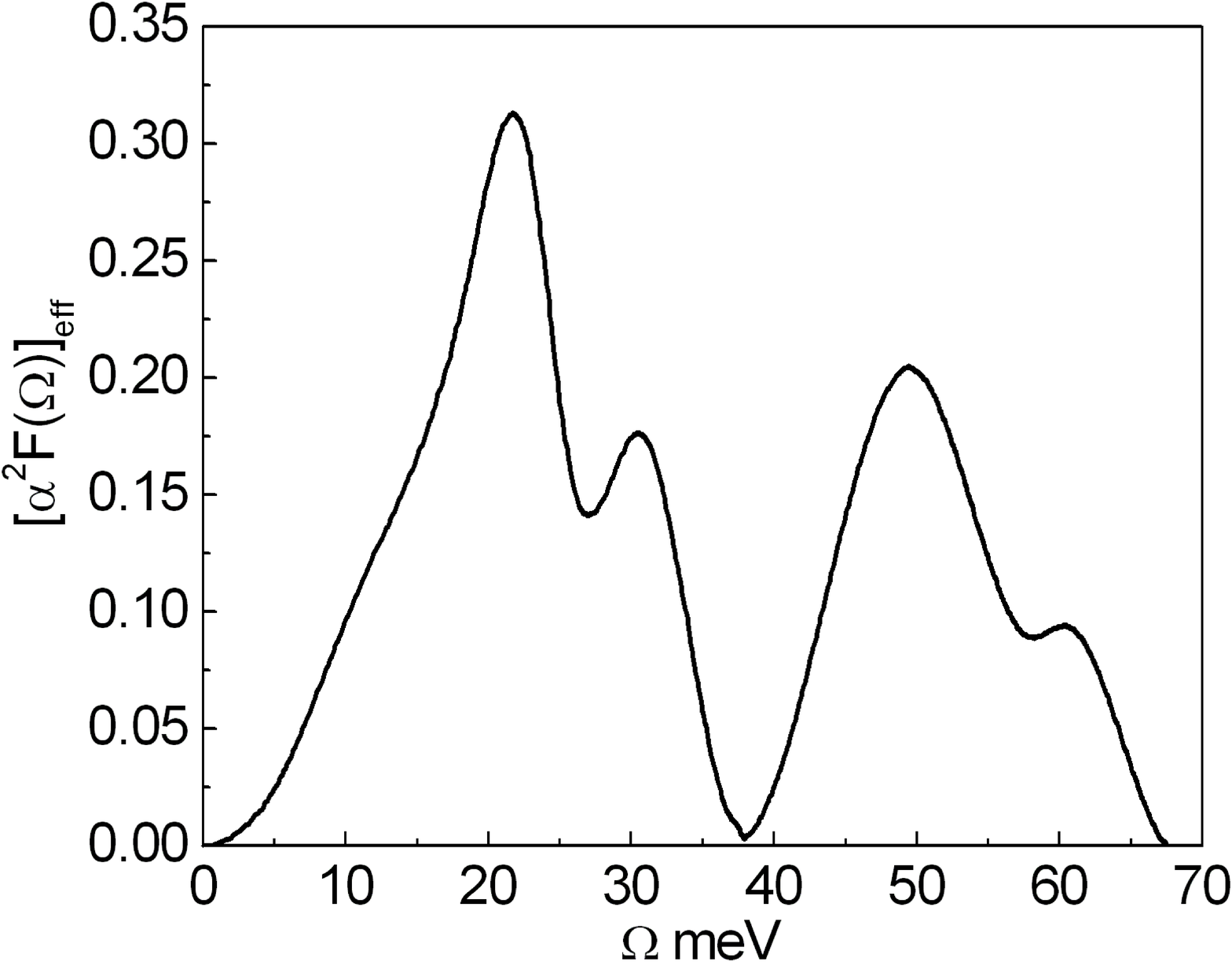}
\caption{\label{f1}
The form of the effective Eliashberg function determined by using the transport function obtained by Gonnelli {\it et al.} \cite{Gonnelli}.}
\end{figure}

In the Eliashberg formalism the coupling between the electron and phonon system is considered with using of the formula:
\begin{equation}
\label{r3}
K\left(\omega_{n}-\omega_{m}\right)\equiv 2\int_0^{\Omega_{\rm{max}}}d\Omega\frac{\left[\alpha^{2}F\left(\Omega\right)\right]_{\rm{eff}}\Omega}
{\left(\omega_n-\omega_m\right)^2+\Omega ^2}.
\end{equation}
The effective Eliashberg function $\left[\alpha^{2}F\left(\Omega\right)\right]_{\rm{eff}}$ for $\rm{YNi_{2}B_{2}C}$ superconductor is shown in Fig.\fig{f1}. On the basis of the presented data, it is easy to see an exceptionally strong coupling that appears between the electron gas and the crystal lattice vibrations of frequency of about $20$ and $50$ meV. The mentioned effect was confirmed with a use of the point-contact spectroscopy measurements in \cite{Naidyuk}. The value of the maximum phonon frequency ($\Omega_{\rm{max}}$) is equal to $67.51$ meV. 

%
%

From the mathematical point of view, the strict analysis of the Eliashberg equations on the imaginary axis, however complicated, is much simpler than on the real axis. In the imaginary domain the arguments of functions $\Delta_{n}$ and $Z_{n}$ take discreet values, thus all problems connected with the numerical analysis of the non-linear integral equations are eliminated \cite{Eliashberg}. On the imaginary axis the Eliashberg equations can be solved in the following way \cite{Szczesniak1}: in the first step one need to define functions $K_{n,m}^{\left(+\right)}$ and $K_{n,m}^{\left(-\right)}$  for quantities $\phi_{n}\equiv\Delta_{n}Z_{n}$ and $Z_{n}$ respectively, where: $K_{n,m}^{\left(\pm\right)}\equiv K\left(\omega_{n}-\omega_{m}\right)\pm K\left(\omega_{n}-\omega_{-m+1}\right)$. Next, the fact that functions $\phi_{n}$ and $Z_{n}$ are symmetric is used. The Eliashberg equations take the form:
\begin{equation}
\label{r4}
\phi_{n}=\sum_{m=1}^{M}J_{1}\left(\omega_{n},\omega_{m},Z_{m},\phi_{m}\right)\phi_{m},
\end{equation}
\begin{equation}
\label{r5}
Z_{n}=\sum_{m=1}^{M}J_{2}\left(\omega_{n},\omega_{m},Z_{m},\phi_{m}\right)Z_{m},
\end{equation}
where:
\begin{equation}
\label{r6}
J_{1}\left(\omega_{n},\omega_{m},Z_{m},\phi_{m}\right)\equiv \frac{\pi}{\beta}\frac{K_{n,m}^{\left(+\right)}}
{\sqrt{\left(Z_{m}\omega_{m}\right)^{2}+\phi^{2}_{m}}}
\end{equation}
and
\begin{equation}
\label{r7}
J_{2}\left(\omega_{n},\omega_{m},Z_{m},\phi_{m}\right)\equiv
\frac{\delta_{n,m}}{Z_{m}}+\frac{\pi}{\beta}\frac{\omega_{m}}{\omega_{n}}\frac{K_{n,m}^{\left(-\right)}}
{\sqrt{\left(Z_{m}\omega_{m}\right)^{2}+\phi^{2}_{m}}}.
\end{equation}
The symbol $\delta_{n,m}$ appearing in Eq. \eq{r7} denotes the Kronecker delta. Let us notice, that the parameter $M$ must be chosen in a such way, that the solutions of the Eliashberg equations for the large values of $n$ and $T=\left[T\right]_{\rm{min}}$ would take their asymptotic form: $\Delta_{n}\simeq 0$ and $Z_{n}\simeq 1$. In the case of $\rm{YNi_{2}B_{2}C}$ it is enough to assume $M=800$. Finally, the Eliashberg set is solved in an iterative way.

\section{The numerical results}

\begin{figure} [ht]
\includegraphics[scale=0.15]{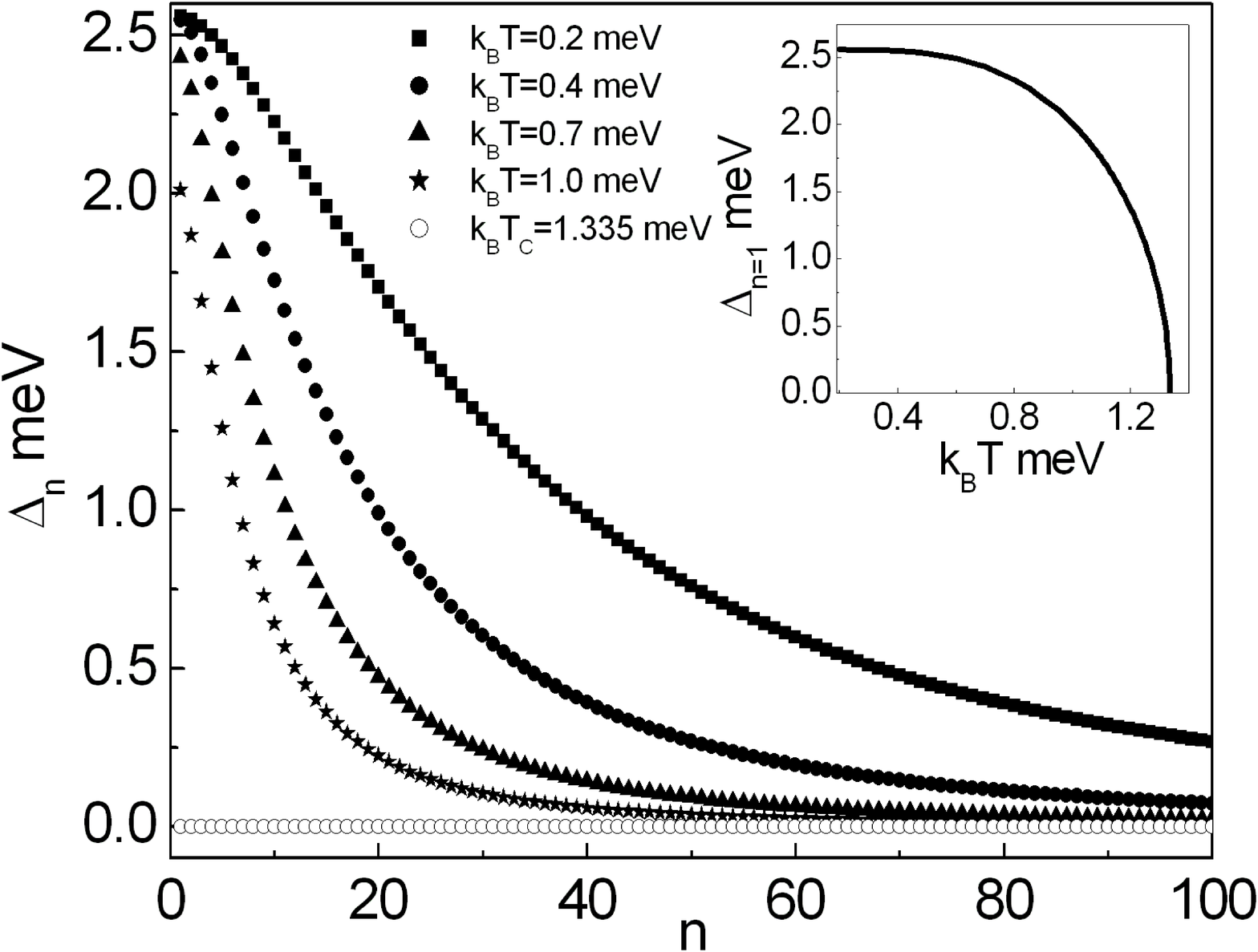}
\caption{\label{f2}
The order parameter function on the imaginary axis for the selected values of the temperature; the first $100$ values of $\Delta_{n}$ are presented. The dependence of the order parameter $\Delta_{n=1}$ on the temperature is plotted in the inset.}
\end{figure}

The Eliashberg equations were solved for the temperature range from $k_{B}\left[T\right]_{\rm{min}}=0.2$ meV to $k_{B}T_{C}=1.335$ meV. The form of the order parameter for the selected values of the temperature is presented in Fig.\fig{f2}. It is easy to notice that together with the decreasing of the temperature, the function $\Delta_{n}$ takes the higher maximum values (always for $n=1$) and it becomes wider. The dependence of the order parameter on the temperature can be traced, in the most convenient way, by plotting the function $\Delta_{n=1}\left(T\right)$; see inset in Fig.\fig{f2}. In the considered case we have taken $350$ accurate numerical values of the order parameter for $n=1$. We notice that the function $\Delta_{n=1}\left(T\right)$ can be fitted by the simple formula:
\begin{equation}
\label{r8}
\Delta_{n=1}\left(T\right)=\Delta_{n=1}\left(0\right)\sqrt{1-\left(\frac{T}{T_{C}}\right)^{\beta}},
\end{equation}
where $\Delta_{n=1}\left(0\right)\equiv \Delta_{n=1}\left(k_{B}T=0.2\quad\rm{meV}\right)=2.559$ meV and $\beta=3.21$.
%
%

\begin{figure} [ht]
\includegraphics[scale=0.15]{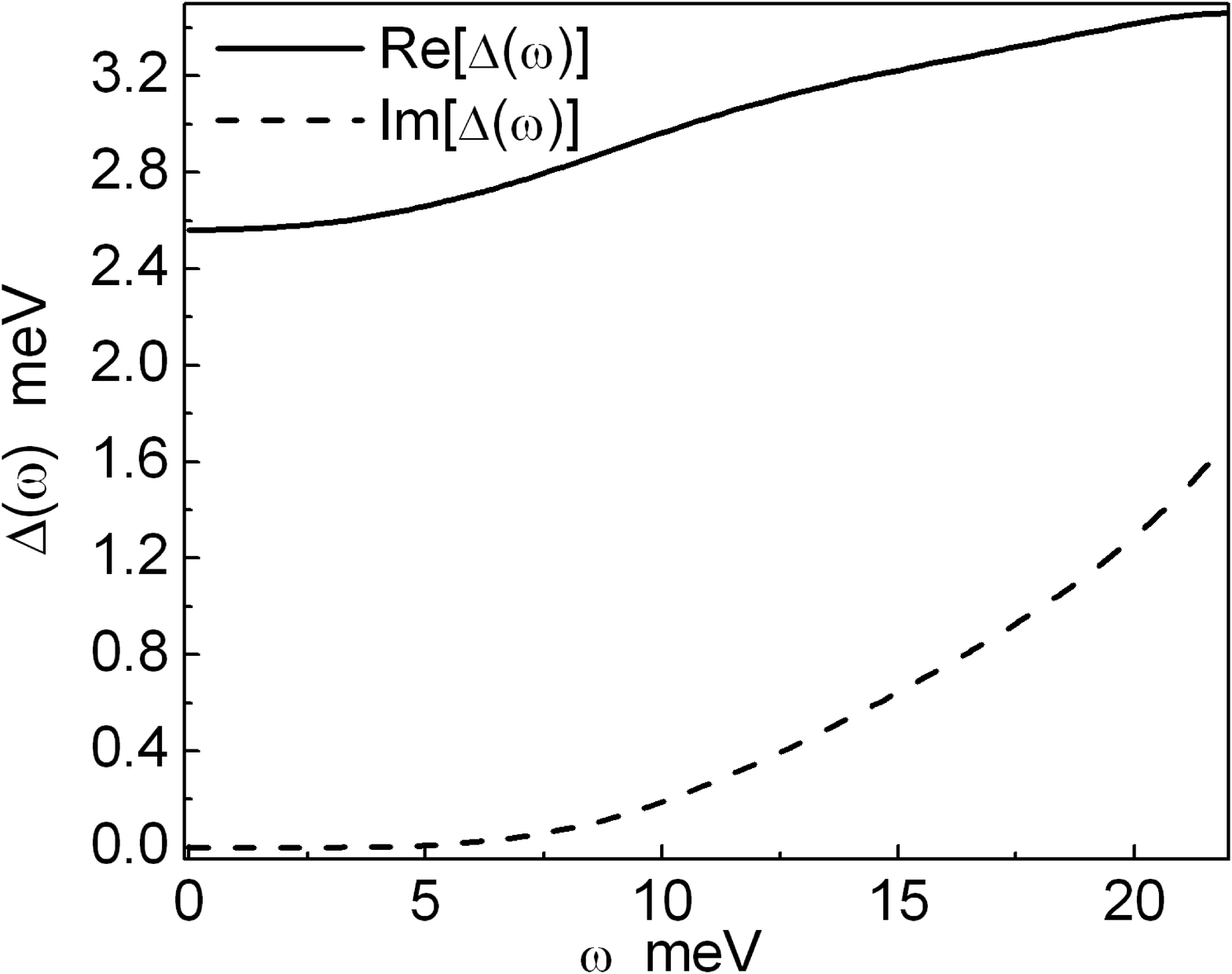}
\caption{\label{f3}
The real and imaginary part of the order parameter on the real axis; $k_{B}T = 0.2$ meV was assumed.}
\end{figure}

The familiarity with the form of the function $\Delta_{n}$ for $k_{B}T = 0.2$ meV enables the calculation of the order parameter value near the temperature of zero Kelvin ($\Delta\left(0\right)$). In particular it was assumed that $\Delta\left(0\right)\simeq\Delta\left(k_{B}T = 0.2\quad\rm{meV}\right)$. In order to achieve that, the order parameter on the imaginary axis needs to be analytically continued on the real axis ($\Delta\left(\omega\right)$), deriving the coefficients $p_{\Delta j}$ and $q_{\Delta j}$ in the expression \cite{Beach}, \cite{Vidberg}:
\begin{equation}
\label{r9}
\Delta\left(\omega\right)=\frac{p_{\Delta 1}+p_{\Delta 2}\omega+...+p_{\Delta r}\omega^{r-1}}
{q_{\Delta 1}+q_{\Delta 2}\omega+...+q_{\Delta r}\omega^{r-1}+\omega^{r}},
\end{equation}
where $r = 400$. The plot of the real (Re) and imaginary (Im) part of the function $\Delta\left(\omega\right)$ is shown in Fig.\fig{f3}. In the last step, the parameter $\Delta\left(0\right)$ should be calculated on the basis of the equation \cite{Carbotte}, \cite{Carbotte1}, \cite{Eliashberg}: $\Delta\left(T\right)={\rm Re}\left[\Delta\left(\omega=\Delta\left(T\right),T\right)\right]$. The value $2.584$ meV was obtained. 

The determination of the parameter $\Delta\left(0\right)$ allows to estimate the dimensionless ratio: $R_{1}\equiv\frac{2\Delta\left(0\right)}{k_{B}T_{C}}$ which, in the BCS theory, is the universal constant of the model and $\left[R_{1}\right]_{\rm{BCS}}=3.53$ \cite{BCS}. In the case of $\rm{YNi_{2}B_{2}C}$, the greater value of $R_{1}$, equal to $3.87$, was obtained. 
With reference to the result predicted by the two-band Eliashberg model, the value of $R_{1}$ estimated in the framework of the one-band model should be interpreted as the resultant quantity. In particular the two-band Eliashberg model, which very well reconstructs the experimental value of $R_{1}$, predicts: $R_{1}=0.71\left[R_{1}\right]_{l}+0.29\left[R_{1}\right]_{s}=3.78$, where $\left[R_{1}\right]_{l}$ and $\left[R_{1}\right]_{s}$ denote the ratios for the large and the small value of the order parameter respectively \cite{Huang}. When comparing the obtained results it is easy to notice, that the one-band model insignificantly overestimates $R_{1}$.   
%
%

\begin{figure} [ht]
\includegraphics[scale=0.15]{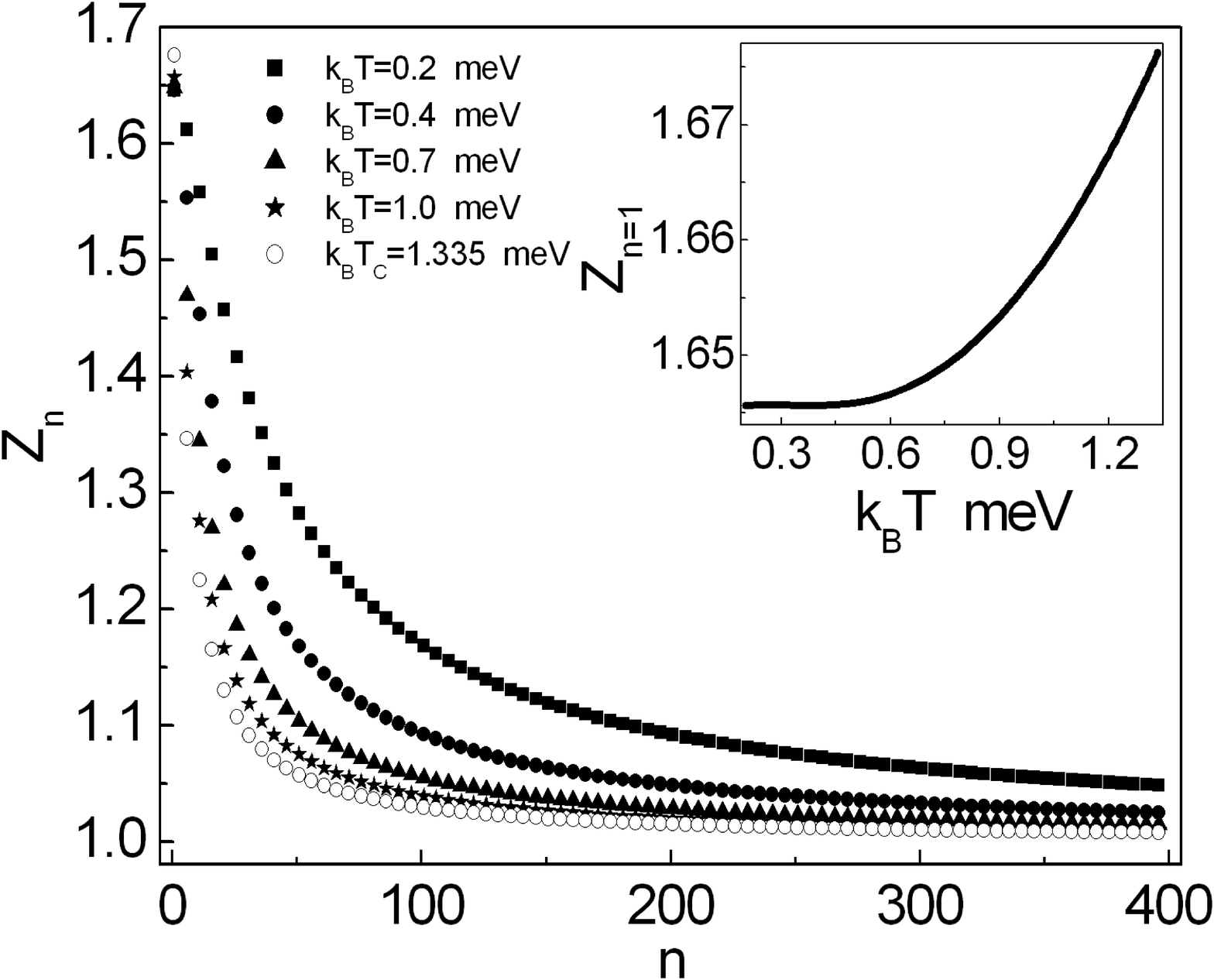}
\caption{\label{f4}
The wave function renormalization factor for the selected values of the temperature; the first $400$ values of $Z_{n}$ are presented. The dependence of the wave function renormalization factor on the temperature for the first Matsubara frequency is plotted in the inset.}
\end{figure}

In Fig.\fig{f4} the dependence of the wave function renormalization factor on the Matsubara frequency for the selected values of the temperature is presented. On the basis of the achieved results it has been stated, that function $Z_{n}$ takes its maximum always for $n=1$. In the framework of the Eliashberg formalism the value $Z_{n=1}$ plays very important role, because it determines the ratio: $Z_{n=1}=m_{e}^{*}/m_{e}$, where $m_{e}^{*}$ denotes the electron band effective mass in the presence of the electron-phonon coupling and $m_{e}$ is the electron band effective mass in absence of the electron-phonon interaction. Basing on the results presented in Figure's \fig{f4} inset, it has been concluded, that $m_{e}^{*}$ takes its maximum value equal to $1.676 m_{e}$ for $T = T_{C}$. Let us mark the fact, that in the case $T = T_{C}$, the electron band effective mass in the presence of the electron-phonon coupling can be calculated with an use of the formula:
\begin{equation}
\label{r10}
m_{e}^{*}=\left(1+\lambda\right)m_{e}.
\end{equation}
On the basis of expression \eq{r10} it has been stated, that the calculated value of $m_{e}^{*}$ is identical with the value obtained using the Eliashberg equations. The above result partially confirms the accurateness of the advanced numerical calculations.

%
%

Next, we have calculated the following ratios:
\begin{equation}
\label{r11}
R_{2}\equiv\frac{\Delta C\left(T_{C}\right)}{C^{N}\left(T_{C}\right)} 
\end{equation}
and
\begin{equation}
\label{r12}
R_{3}\equiv\frac{T_{C}C^{N}\left(T_{C}\right)}{H_{C}^{2}\left(0\right)},
\end{equation}
where $\Delta C\equiv C^{S}-C^{N}$ stands for the difference between specific heat of the superconducting and normal state. The dependence of $\Delta C$ on the temperature is being determined with a use of an expression:
\begin{equation}
\label{r13}
\frac{\Delta C\left(T\right)}{k_{B}\rho\left(0\right)}
=-\frac{1}{\beta}\frac{d^{2}\left[\Delta F/\rho\left(0\right)\right]}
{d\left(k_{B}T\right)^{2}}.
\end{equation}
In Eq. \eq{r13} the symbol $\Delta F$ denotes the difference between the free energy of the superconducting and normal state; $\rho\left(0\right)$ is the value of the electron density of states at the Fermi level. 

On the other hand, the specific heat in the normal state can be calculated using the formula: $\frac{C^{N}\left(T\right)}{ k_{B}\rho\left(0\right)}=\frac{\gamma}{\beta}$, where $\gamma\equiv\frac{2}{3}\pi^{2}\left(1+\lambda\right)$. 
Finally, the thermodynamic critical field ($H_{C}$) should be estimated in accordance with the expression: 
$\frac{H_{C}}{\sqrt{\rho\left(0\right)}}=\sqrt{-8\pi\left[\Delta F/\rho\left(0\right)\right]}$.
 
Having an open solutions of the Eliashberg equations, $\Delta F$ is determined directly from \cite{Bardeen}:
\begin{eqnarray}
\label{r14}
\frac{\Delta F}{\rho\left(0\right)}&=&-\frac{2\pi}{\beta}\sum_{n=1}^{M}
\left(\sqrt{\omega^{2}_{n}+\Delta^{2}_{n}}- \left|\omega_{n}\right|\right)\\ \nonumber
&\times&(Z^{{\rm S}}_{n}-Z^{N}_{n}\frac{\left|\omega_{n}\right|}
{\sqrt{\omega^{2}_{n}+\Delta^{2}_{n}}}),  
\end{eqnarray}  
where symbols $Z^{{\rm S}}_{n}$ and $Z^{{\rm N}}_{n}$ denote the wave function renormalization factor for the superconducting state and the normal state respectively.

%
\begin{figure} [ht]
\includegraphics[scale=0.15]{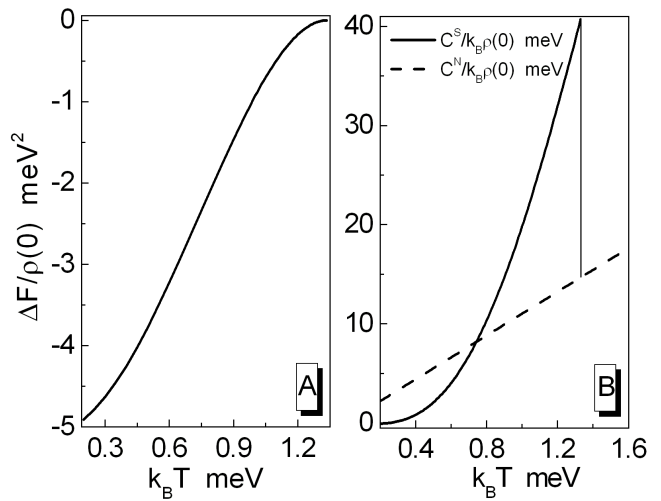}
\caption{\label{f5}
(A) The dependence of the free energy difference between the superconducting and normal state on the temperature. 
(B) The specific heat of the superconducting and normal state as a function of the temperature.} 
\end{figure}
%

The determined form of $\Delta F\left(T\right)$ is presented in Fig.\fig{f5} (A), while in Fig.\fig{f5} (B) we are shown the dependence of the specific heat on the temperature for the superconducting and normal state. The characteristic jump of the specific heat that appears at the critical temperature was marked with a vertical line. We notice that the specific heat of the superconducting state was obtained on the basis of the 350 accurate values of $\Delta F/\rho\left(0\right)$. 

On the basis of the determined thermodynamic functions, the parameters $R_{2}$ and $R_{3}$ were calculated. In the case of $\rm{YNi_{2}B_{2}C}$ these coefficients take following values: $1.79$ and $0.159$. Let us notice, that $R_{2}$ and $R_{3}$ are the universal constants of the BCS model and  $\left[R_{2}\right]_{\rm{BCS}}=1.43$ and $\left[R_{3}\right]_{\rm{BCS}}=0.168$ \cite{BCS}. When basing on the above results one can clearly see, that for $\rm{YNi_{2}B_{2}C}$ superconductor the value of ratio $R_{2}$ significantly deviates from the BCS prediction.

The experimental ratios $R_{2}$ and $R_{3}$ were determined in \cite{Michor}. The following results were obtained: $\left[R_{2}\right]_{\rm{exp}}=1.77$ and $\left[R_{3}\right]_{\rm{exp}}=0.160$. When comparing our theoretical values of $R_{2}$ and $R_{3}$ with the experimental ones, it can be easily noticed that the one-band Eliashberg model properly describes the experimental data.

\section{Concluding Remarks}

In the framework of the one-band Eliashberg model the selected thermodynamic properties of  $\rm{YNi_{2}B_{2}C}$ superconductor was calculated. In particular, the fundamental ratios $R_{1}\sim R_{3}$ were determined. It has been stated that, in the case of $R_{1}$ the one-band Eliashberg model predicts a value non-significantly higher than the value determined by the two-band model, which determines $R_{1}$ with a very high accuracy. The result above is connected with the simplified description of the superconducting phase in the one-band model in comparison with the two-band model; we have taken the one effective Eliashberg function instead of the four Eliashberg functions and the four elements of the Coulomb pseudopotential matrix.

In the cases of the two remain ratios - the agreement between predictions of the one-band Eliashberg model and the experimental data is very good. It is worth to notice that, in the opposition to $R_{1}$ and $R_{3}$, the value of $R_{2}$ significantly deviates from the prediction of the BCS model.

In the last part of the summary let us turn a reader's attention toward the problem of the correct determination of the upper critical field. When basing on the papers cited in the introduction one can suppose with a large probability, that the exact form of the function $H_{C2}\left(T\right)$ can be reproduced only in the framework of the two-band theory. Let us remind that the two-band approach was successfully used by us for description of the thermodynamic properties of the superconducting state inducing in $\rm{MgB_{2}}$ \cite{MgB2}. However, in the case of $\rm{YNi_{2}B_{2}C}$ discussed problem is far more complicated because the appropriate Eliashberg functions have not been calculated in the branch press (only electron-phonon coupling constants are being used in the existing two-band approach). At present this issue is being intensively studied by us with a use of the {\it ab initio}  approach \cite{AbInitio}.      

\begin{acknowledgments}
The authors wish to thank Prof. K. Dzili{\'n}ski for the creation of the excellent working conditions and providing the financial support. Some computational resources have been provided by the RSC Computing Center.
\end{acknowledgments}
%

%
\end{document}